\documentclass[PRL,preprintnumbers,floatfix,aps,notitlepage,nofootinbib,twocolumn,showpacs,amssymb]{revtex4-2}
\usepackage{amsmath,amsfonts,bm}
\usepackage{graphicx}
\usepackage{cancel}
\usepackage[colorlinks, linkcolor={red},citecolor={blue}]{hyperref}

\begin{document}

\title{Energy exchange between relativistic fluids: the polytropic case}
\author{J. Ovalle$^{ab}$}
\email[]{jorge.ovalle@physics.slu.cz (corresponding author)}
\affiliation{$^a$Research Centre for Theoretical Physics and Astrophysics,
	Institute of Physics, Silesian University in Opava, CZ-746 01 Opava, Czech Republic.\\ \\
	$^b$Departamento de F\'isica, Facultad de Ciencias B\'asicas,
	Universidad de Antofagasta, Chile.}

\author{E. Contreras}
\email{econtreras@usfq.edu.ec}
\affiliation{Departamento de F\'isica, Colegio de Ciencias e Ingenier\'ia,
	Universidad San Francisco de Quito, Quito, Ecuador.
}

	\author{Z. Stuchlik}
	\email[]{zdenek.stuchlik@physics.slu.cz}
	\affiliation{Research Centre for Theoretical Physics and Astrophysics,
		Institute of Physics, Silesian University in Opava, CZ 746 01 Opava, Czech Republic.}

\begin{abstract}
	We present a simple, analytic and straightforward method to elucidate the effects produced by polytropic fluids on any other gravitational source, no matter its nature, for static and spherically symmetric spacetimes. As a direct application, we study the interaction between polytropes and perfect fluids coexisting inside a self-gravitating stellar object.
		\end{abstract} 
\maketitle
%
%

\section{Introduction}
The study of self-gravitating systems is of great importance in the context of general relativity. In particular, elucidating what happens inside compact stellar distributions is extremely important, especially if we want to gain a good understanding of gravitational collapse. Presumably, the interior of a stellar structure is a complex physical system, formed by fluids of different natures, which surely interact with each other in a non-trivial way. Although it is true that we can always group fluids of different nature into a single energy-momentum tensor $\tilde{T}_{\mu\nu}$, namely,
\begin{equation}
	\label{1}
\tilde{T}_{\mu\nu}=T^{1}_{\mu\nu}+...+T^{n}_{\mu\nu}\ ,
\end{equation}
as indeed it is required by Einstein's field equations, it is no less true that an oversimplification of these systems could jeopardize an adequate description of them. A clear example of this occurs when we consider complex stellar distributions (being electrically charged, with dissipative effects, anisotropic, etc.) and impose a fairly simple equation of state to describe the system as a whole. This could mean an oversimplification whose main motivation is far from being physical: a reduction of the degrees of freedom of the system that makes it more manageable. It is true that in many cases this strategy produces great results, such as exact and physically reasonable solutions, but perhaps we are paying a high price without being aware of it, such as a rather idealized description of a system whose physical nature is intrinsically complex.
\par
Regarding the above, it would be very useful to see how relevant is the role that each fluid, represented by their respective energy-momentum tensor $T^{i}_{\mu\nu}$ in~\eqref{1}, plays on a self-gravitating system, as well as how these gravitational sources interact with each other. This would allow, for instance, detecting which source dominates over the others, and consequently rule out any equation of state incompatible with the dominant source.  Conceptually, achieving this in general relativity should be extremely difficult, given the nonlinear nature of the theory. However, since the Gravitational Decoupling approach (GD)~\cite{Ovalle:2017fgl,Ovalle:2019qyi} is precisely designed for coupling/decoupling gravitational sources in general relativity, we will see that, indeed, it is possible to elucidate the role played by each gravitational source, without resorting to any numerical protocol or perturbation scheme, as explained in the next paragraph.
\par 
In particular, if in Eq.~\eqref{1}  we consider two arbitrary sources $\{T_{\mu\nu},\,\theta_{\mu\nu}\}$, then the contracted Bianchi identities  yields $\nabla_\mu\,T^{\mu\nu}+\nabla_\mu\,\theta^{\mu\nu}=0$. This has two possible solutions, namely, 
\begin{eqnarray}
	\label{b1}
	&&\nabla_\mu\,T^{\mu}_{\ \nu}=\nabla_\mu\theta^{\mu}_{\ \nu}=0\nonumber\ ,
	\\
	\label{b2}
	&&\nabla_\mu\,T^{\mu}_{\ \nu}=-\nabla_\mu\theta^{\mu}_{\ \nu}= \textbf{?}\nonumber
\end{eqnarray}
The first solution indicates that each source is covariantly conserved, and therefore the interaction between them is purely gravitational. The second option, more interesting and much more realistic,\footnote{If both fluids coexist in a certain region of spacetime, as in fact occurs within a self-gravitating object.} indicates an exchange of energy between these sources that, in principle, would be impossible to quantify or at least describe in some detail. The reason for this is that the Bianchi identities do not introduce additional information beyond Einstein's equations. They are identities, and therefore are trivially satisfied. However, the GD is precisely the scheme that circumvents the intrinsic triviality of Bianchi identities, and it is what allows to elucidate, in detail, the interaction between both gravitational sources [see further Eq.~\eqref{con22}]. 
\par
In particular, and motivated by the interest that they have generated in recent years, in this article we will choose a polytrope as one of the gravitational sources to determine its effects on any other generic fluid, regardless of its nature.
\par
The paper is organised as follows:
in Section~\ref{sec2}, we first review the fundamentals of the GD approach 
to a spherically symmetric system containing two generic sources;
in Section~\ref{sec3}, we choose a polytropic fluid to study its effects on a generic gravitational source, 
and we introduce a systematic and direct procedure to elucidate these effects; in Section~\ref{sec4}, we implement the strategy developed in Section~\ref{sec3} for the case of a perfect fluid; finally, we summarize our conclusions in Section~\ref{con}.
\section{Gravitational Decoupling}
\label{sec2}
In this Section, we briefly review the GD for spherically symmetric gravitational systems
described in detail in Ref.~\cite{Ovalle:2019qyi}. For the axially symmetric case, see Ref.~\cite{Contreras:2021yxe}.
The gravitational decoupling approach and its simplest version~\cite{Ovalle:2017fgl}, based in the Minimal Geometric Deformation (MGD)~\cite{Ovalle:2020fuo,daRocha:2017cxu,daRocha:2017lqj,Fernandes-Silva:2017nec,Casadio:2017sze,
	Fernandes-Silva:2018abr,Contreras:2018vph,Contreras:2018gzd,Contreras:2018nfg,
	Panotopoulos:2018law,daRocha:2019pla,Heras:2019ibr,Rincon:2019jal,
	daRocha:2020rda,Contreras:2020fcj,Arias:2020hwz,daRocha:2020jdj,Tello-Ortiz:2020euy,
	daRocha:2020gee,Meert:2020sqv,Tello-Ortiz:2021kxg,Maurya:2021huv,Azmat:2021kmv,Maurya:2021zvb}, are attractive
for many reasons (for an incomplete list of references, see~\cite{Ovalle:2017wqi,
	Gabbanelli:2018bhs,Heras:2018cpz,Estrada:2018zbh,Sharif:2018toc,
	Sharif:2018pzr,Morales:2018urp,Estrada:2018vrl,Sharif:2018tiz,Ovalle:2018umz,
	Contreras:2019fbk,Contreras:2019iwm,Contreras:2019mhf,
	Gabbanelli:2019txr,Estrada:2019aeh,Ovalle:2019lbs,Ovalle:2019lbs,
	Hensh:2019rtb,Cedeno:2019qkf,Torres:2019mee,
	Casadio:2019usg,Singh:2019ktp,Maurya:2019noq,Sharif:2019mjn,Singh:2019ktp,
	Abellan:2020wjw,Sharif:2020vvk,Tello-Ortiz:2020ydf,
	Maurya:2020rny,Rincon:2020izv,Sharif:2020arn,Maurya:2020gjw,Zubair:2020lna,Sharif:2020rlt,Ovalle:2020kpd,Estrada:2020ptc,Maurya:2020djz,Maurya:2021aio,Azmat:2021qig,Islam:2021dyk,Afrin:2021imp,Ovalle:2021jzf,Ama-Tul-Mughani:2021ewd,Sharif:2021emv,daRocha:2021aww,Maurya:2021qye,Carrasco-Hidalgo:2021dyg,Sultana:2021cvq,daRocha:2021sqd,Maurya:2021yhc,Omwoyo:2021uah,Afrin:2021ggx,Meert:2021khi}. Among them we can mention i) the coupling of gravitational sources, which allows for extending known solutions of the Einstein field equations into 
more complex domains; ii) the decoupling of gravitational sources, which is used to systematically reduce (decouple) a complex energy-momentum
tensor $T_{\mu\nu}$ into simpler components; iii) to find solutions in gravitational theories beyond Einstein's; iv) to generate rotating hairy black hole solutions, among many others applications. 

\par
Let us consider the Einstein field equations~\footnote{We use units with $c=1$ and $\kappa\,=8\,\pi\,G_{\rm N}$,
	where $G_{\rm N}$ is Newton's constant.}
\begin{equation}
	\label{corr2}
	G_{\mu\nu}
	\equiv
	R_{\mu\nu}-\frac{1}{2}\,R\, g_{\mu\nu}
	=
	\kappa\,\,\tilde{T}_{\mu\nu}
	\ ,
\end{equation}
with a total energy-momentum tensor given by,
\begin{equation}
	\label{emt}
	\tilde{T}_{\mu\nu}
	=
	T^{\rm}_{\mu\nu}
	+
	\theta_{\mu\nu}
	\ ,
\end{equation}
where $T_{\mu\nu}$ is usually associated with some already known solution, 
whereas $\theta_{\mu\nu}$ may contain new fields or even be related with a new gravitational sector not described by general relativity.
As a consequence of Bianchi identity, the total source must be covariantly conserved,
\begin{equation}
	\nabla_\mu\,\tilde{T}^{\mu\nu}=0
	\ .
	\label{dT0}
\end{equation} 
For spherically symmetric and static systems, we can write the metric $g_{\mu\nu}$ as
\begin{equation}
	ds^{2}
	=
	e^{\nu (r)}\,dt^{2}-e^{\lambda (r)}\,dr^{2}
	-r^{2}\,d\Omega^2
	\ ,
	\label{metric}
\end{equation}
where $\nu =\nu (r)$ and $\lambda =\lambda (r)$ are functions of the areal
radius $r$ only and $d\Omega^2=d\theta^{2}+\sin ^{2}\theta \,d\phi ^{2}$.
The Einstein equations~(\ref{corr2}) then read
\begin{eqnarray}
	\label{ec1}
	\kappa\,\!
	\left(
	T_0^{\ 0}+\theta_0^{\ 0}
	\right)
	&=&
	\frac 1{r^2}
	-
	e^{-\lambda }\left( \frac1{r^2}-\frac{\lambda'}r\right)
	\\
	\label{ec2}
	\kappa\,\!
	\left(T_1^{\ 1}+\theta_1^{\ 1}\right)
	&=&
	\frac 1{r^2}
	-
	e^{-\lambda }\left( \frac 1{r^2}+\frac{\nu'}r\right)
	\\
	\label{ec3}
	\kappa\,\!
	\left(T_2^{\ 2}+\theta_2^{\ 2}\right)
	&=&
	-\frac {e^{-\lambda }}{4}
	\left(2\nu''+\nu'^2-\lambda'\nu'
	+2\,\frac{\nu'-\lambda'}r\right)
	\ ,
\end{eqnarray}
where $f'\equiv \partial_r f$ and $\tilde{T}_3^{{\ 3}}=\tilde{T}_2^{\ 2}$ due to the spherical symmetry.
By simple inspection, we can identify in Eqs.~\eqref{ec1}-\eqref{ec3} an effective density  
\begin{equation}
	\tilde{\rho}
	=
	T_0^{\ 0}
	+
	\theta_0^{\ 0}=\rho+{\cal E}
	\ ,
	\label{efecden}
\end{equation}
an effective radial pressure
\begin{equation}
	\tilde{p}_{r}
	=
	-T_1^{\ 1}
	-\theta_1^{\ 1}=p_r+{\cal P}_r
	\ ,
	\label{efecprera}
\end{equation}
and an effective tangential pressure
\begin{equation}
	\tilde{p}_{t}
	=
	-T_2^{\ 2}
	-\theta_2^{\ 2}=p_t+{\cal P}_t
	\ ,
	\label{efecpretan}
\end{equation} 
where clearly we have 
\begin{eqnarray}
	&&T_\mu^{\,\,\nu}=diag[\rho,\,-p_r,\,-p_t,\,-p_t]\ ,\\
\nonumber	\\
	&&\theta_\mu^{\,\,\nu}=diag[{\cal E},\,-{\cal P}_r,\,-{\cal P}_t,\,-{\cal P}_t]\ .
\end{eqnarray}
In general, the anisotropy 
\begin{equation}
	\label{anisotropy}
	\Pi
	\equiv
	\tilde{p}_{t}-\tilde{p}_{r}
\end{equation}
does not vanish and the system of Eqs.~(\ref{ec1})-(\ref{ec3}) may be
treated as an anisotropic fluid.
\par
We next consider a solution to the Eqs.~\eqref{corr2} for the seed source $T_{\mu\nu}$
alone, that is, 
\begin{equation}
	\label{alone}
	\tilde{T}_{\mu\nu}
	=
	T^{\rm}_{\mu\nu}
	+
	\cancelto{0}\theta_{\mu\nu}
	\ ,
\end{equation}
which we write as
\begin{equation}
	ds^{2}
	=
	e^{\xi (r)}\,dt^{2}
	-e^{\mu (r)}\,dr^{2}
	-
	r^{2}\,d\Omega^2
	\ ,
	\label{pfmetric}
\end{equation}
where 
\begin{equation}
	\label{standardGR}
	e^{-\mu(r)}
	\equiv
	1-\frac{\kappa\,}{r}\int_0^r x^2\,T_0^{\, 0}(x)\, dx
	=
	1-\frac{2\,m(r)}{r}
\end{equation}
is the standard general relativity expression containing the Misner-Sharp mass function $m=m(r)$.
The consequences of adding the source $\theta_{\mu\nu}$ can be seen in the geometric deformation of the metric~\eqref{pfmetric}, namely\footnote{usually we write $\alpha\,g$ and $\alpha\,f$, with $\alpha$ a parameter introduced to keep track of these deformations. Here we dispense with it for simplicity. }
\begin{eqnarray}
	\label{gd1}
	\xi 
	&\rightarrow &
	\nu\,=\,\xi+g
	\\
	\label{gd2}
	e^{-\mu} 
	&\rightarrow &
	e^{-\lambda}=e^{-\mu}+f
	\ , 
\end{eqnarray}
where $f$ and $g$ are respectively the geometric deformations for the radial and temporal metric
components. We emphasize that the expressions in Eqs.~\eqref{gd1} and~\eqref{gd2}
	are not a coordinate transformation. They just represent the change in the spacetime
	geometry~\eqref{pfmetric} generated by a physical source with energy-momentum tensor
	$\theta_{\mu\nu}$.
\par
By means of Eqs.~(\ref{gd1}) and (\ref{gd2}), the Einstein equations~(\ref{ec1})-(\ref{ec3})
are separated in two sets:
A) one is given by the standard Einstein field equations with the energy-momentum tensor $T_{\mu\nu}$,
that is
\begin{eqnarray}
	\label{ec1pf}
	&&
	\kappa\,\rho
	=\frac 1{r^2}
	-
	e^{-\mu }\left( \frac1{r^2}-\frac{\mu'}r\right)\ ,
	\\
	&&
	\label{ec2pf}
	\kappa\,
	\,p_r
	=
	-\frac 1{r^2}
	+
	e^{-\mu}\left( \frac 1{r^2}+\frac{\xi'}r\right)\ ,
	\\
	&&
	\label{ec3pf}
	\kappa\,
	\strut\displaystyle
	\,p_t
	=
	\frac {e^{-\mu }}{4}
	\left(2\xi''+\xi'^2-\mu'\xi'
	+2\,\frac{\xi'-\mu'}r\right)
	\ ,
\end{eqnarray}
which is assumed to be solved by the metric~(\ref{pfmetric});
B) the second set contains the source $\theta_{\mu\nu}$ and reads
\begin{eqnarray}
	\label{ec1d}
	\kappa\,{\cal E}
	&=&
	-\frac{f}{r^2}
	-\frac{f'}{r}\ ,
	\\
	\label{ec2d}
	\kappa\,{\cal P}_r
	-Z_1
	&=&
	f\left(\frac{1}{r^2}+\frac{\nu'}{r}\right)
	\\
	\label{ec3d}
	\kappa\,{\cal P}_t
	-Z_2
	&=&
	\frac{f}{4}\left(2\,\nu''+\nu'^2+2\frac{\nu'}{r}\right)
	\nonumber
	\\
	&&
	\frac{f'}{4}\left(\nu'+\frac{2}{r}\right)
	\ ,
\end{eqnarray}
where 
\begin{eqnarray}
	\label{Z1}
	Z_1
	&=&
	\frac{e^{-\mu}\,g'}{r}
	\\
	\label{Z2}
	4\,Z_2&=&e^{-\mu}\left(2g''+g'^2+\frac{2\,g'}{r}+2\xi'\,g'-\mu'g'\right)
	\ .
\end{eqnarray}
Of course the tensor $\theta_{\mu\nu}$ vanishes
when the deformations vanish ($f=g=0$). We see that for the particular case $g=0$, Eqs.~\eqref{ec1d}-\eqref{ec3d} reduce to the simpler
``quasi-Einstein" system of the MGD of Ref.~\cite{Ovalle:2017fgl},
in which $f$ is only determined by $\theta_{\mu\nu}$ and the undeformed metric~\eqref{pfmetric}. Also, notice that the set~\eqref{ec1d}-\eqref{ec3d} contains $\{\xi,\,\mu\}$, and therefore is not independent of~\eqref{ec1pf}-\eqref{ec3pf}. This of course makes sense since both systems represent a simplified version of a more complex whole, described by Eqs.~\eqref{ec1}-\eqref{ec3}. 
\par
Now let us see the conservation equation~\eqref{dT0}, which reads
\begin{eqnarray}
	\label{con111}
	&&\left[\left({T}_1^{\ 1}\right)'
	-
	\frac{\xi'}{2}\left({T}_0^{\ 0}-{T}_1^{\ 1}\right)
	-
	\frac{2}{r}\left({T}_2^{\ 2}-{T}_1^{\ 1}\right)\right]
	\nonumber \\
	&&-\frac{g'}{2}\left({T}_0^{\ 0}-{T}_1^{\ 1}\right)
	\nonumber\\
	&&+\left({\theta}_1^{\ 1}\right)'
	-
	\frac{\nu'}{2}\left({\theta}_0^{\ 0}-{\theta}_1^{\ 1}\right)
	-
	\frac{2}{r}\left({\theta}_2^{\ 2}-{\theta}_1^{\ 1}\right)
	=
	0\ ,
\end{eqnarray}
The bracket represents the divergence of $T_{\mu\nu}$ computed with the
covariant derivative $\nabla^{(\xi,\mu)}$ for the metric~(\ref{pfmetric}),
and is a linear combination of the Einstein field equations~\eqref{ec1pf}-\eqref{ec3pf}.
Since the Einstein tensor ${G}^{(\xi,\mu)}_{\mu\nu}$ for the metric~(\ref{pfmetric}) satisfies its respective
Bianchi identity, the momentum tensor $T_{\mu \nu }$ is conserved in this geometry,
\begin{equation}
	\label{pfcon}
	\nabla^{(\xi,\mu)}_\sigma\,T^{\sigma}_{\ \nu}=0
	\ .
\end{equation}
Notice that 
\begin{equation}
	\label{divs}
	\nabla_\sigma\,T^{\sigma}_{\ \nu}
	=
	\nabla^{(\xi,\mu)}_\sigma\,T^{\sigma}_{\ \nu}
	-
	\frac{g'}{2}\left({T}_0^{\ 0}-{T}_1^{\ 1}\right)\delta^1_\nu
	\ ,
\end{equation}
where the divergence in the left-hand side is calculated with the deformed metric in Eq.~\eqref{metric}. Finally, Eq.~\eqref{con111} becomes
\begin{eqnarray}
	\label{con22}
		\nabla_\sigma\,T^{\sigma}_{\ \nu}=-\nabla_\sigma\theta^{\sigma}_{\ \nu}
	=
	-
	\frac{g'}{2}\left({T}_0^{\ 0}-{T}_1^{\ 1}\right)\delta^{\sigma}_{\ \nu}
	\ ,
\end{eqnarray}
which is also a linear combination of the ``quasi-Einstein'' field equations~\eqref{ec1d}-\eqref{ec3d} 
for the source $\theta_{\mu\nu}$.
We therefore conclude that the two sources $T_{\mu\nu}$ and $\theta_{\mu\nu}$ can be successfully decoupled
by means of the GD. This result is particularly remarkable since it is exact, without requiring any perturbative expansion in   
$f$ or $g$~\cite{Ovalle:2020fuo}.
\par
Finally, in order to be as self-contained as possible, and to clarify the reader any potential confusion about what we developed between Eqs.~\eqref{corr2}-\eqref{con22}, we next describe the intrinsic relationship between gravitational decoupling and energy exchange for coupled/decoupled relativistic fluids.
\begin{enumerate}
	\item 
We first start by considering a gravitational source $T_{\mu\nu}$. According to general relativity, the equation of motions for this source are Einstein's equations, displayed in Eqs.~\eqref{ec1pf}-\eqref{ec3pf}, whose metric contains the geometric functions $\{\xi,\,\mu\}$.

\item 
Then we couple $T_{\mu\nu}$ with a second source $\theta_{\mu\nu}$ by
\[
T_{\mu\nu}\rightarrow\,T_{\mu\nu}+\theta_{\mu\nu}\ .
\]
This coupling changes the original spacetime geometry $\{\xi,\,\mu\}\rightarrow\{\nu,\,\lambda\}$, as specified by Eqs.~\eqref{gd1} and~\eqref{gd2}.

\item 
The new spacetime geometry $\{\nu,\,\lambda\}$, associated with the total source $	\tilde{T}_{\mu\nu}
=
T^{\rm}_{\mu\nu}
+
\theta_{\mu\nu}$, satisfies Einstein's equations~\eqref{ec1}-\eqref{ec3} if and only if the source $\theta_{\mu\nu}$ and its geometric functions $\{g,\,f\}$ satisfy the equation of motions~\eqref{ec1d}-\eqref{ec3d}.
\begin{figure*}[ht!]
	\centering
	\includegraphics[width=0.33\textwidth]{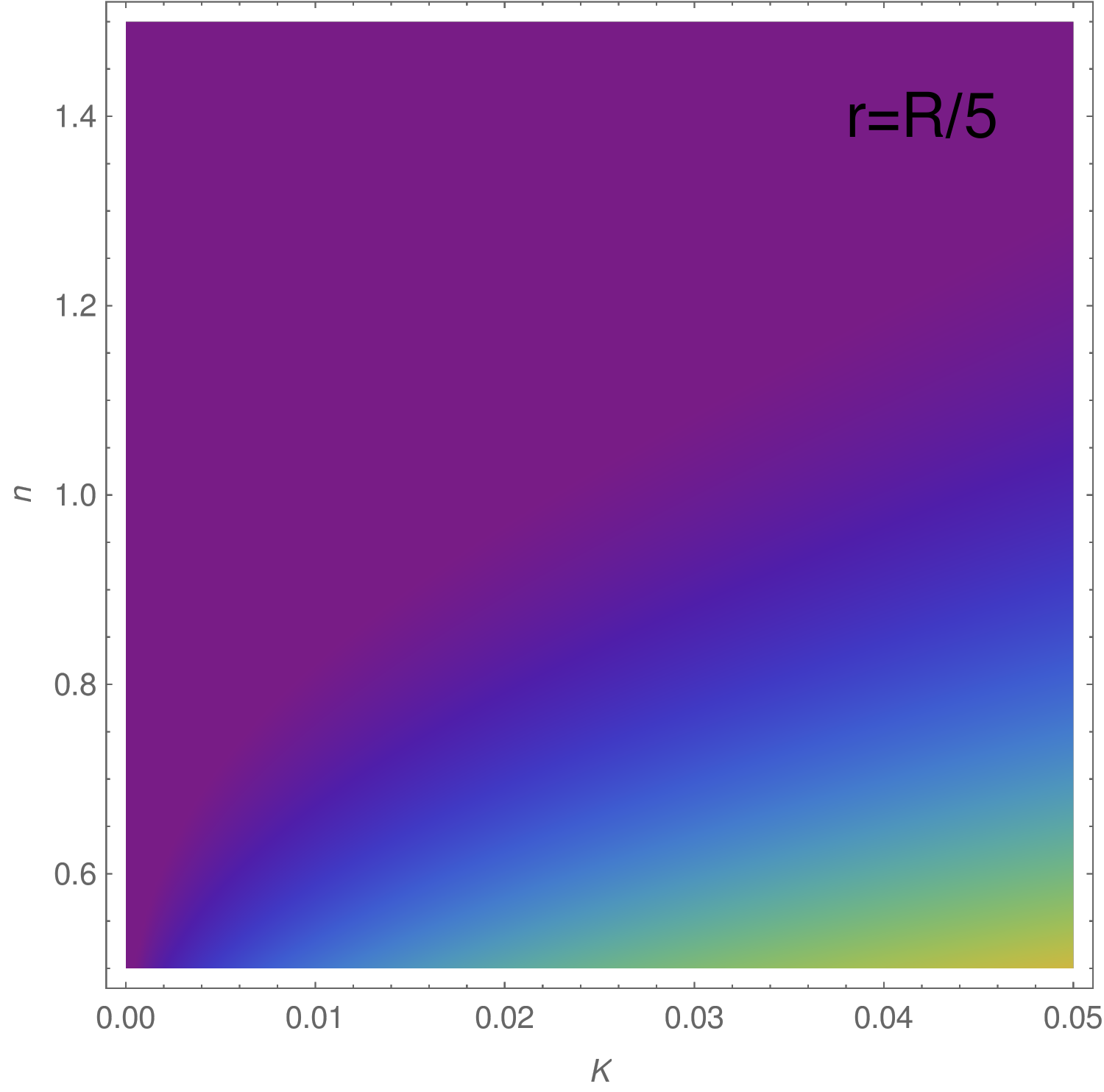} \
	\includegraphics[width=0.33\textwidth]{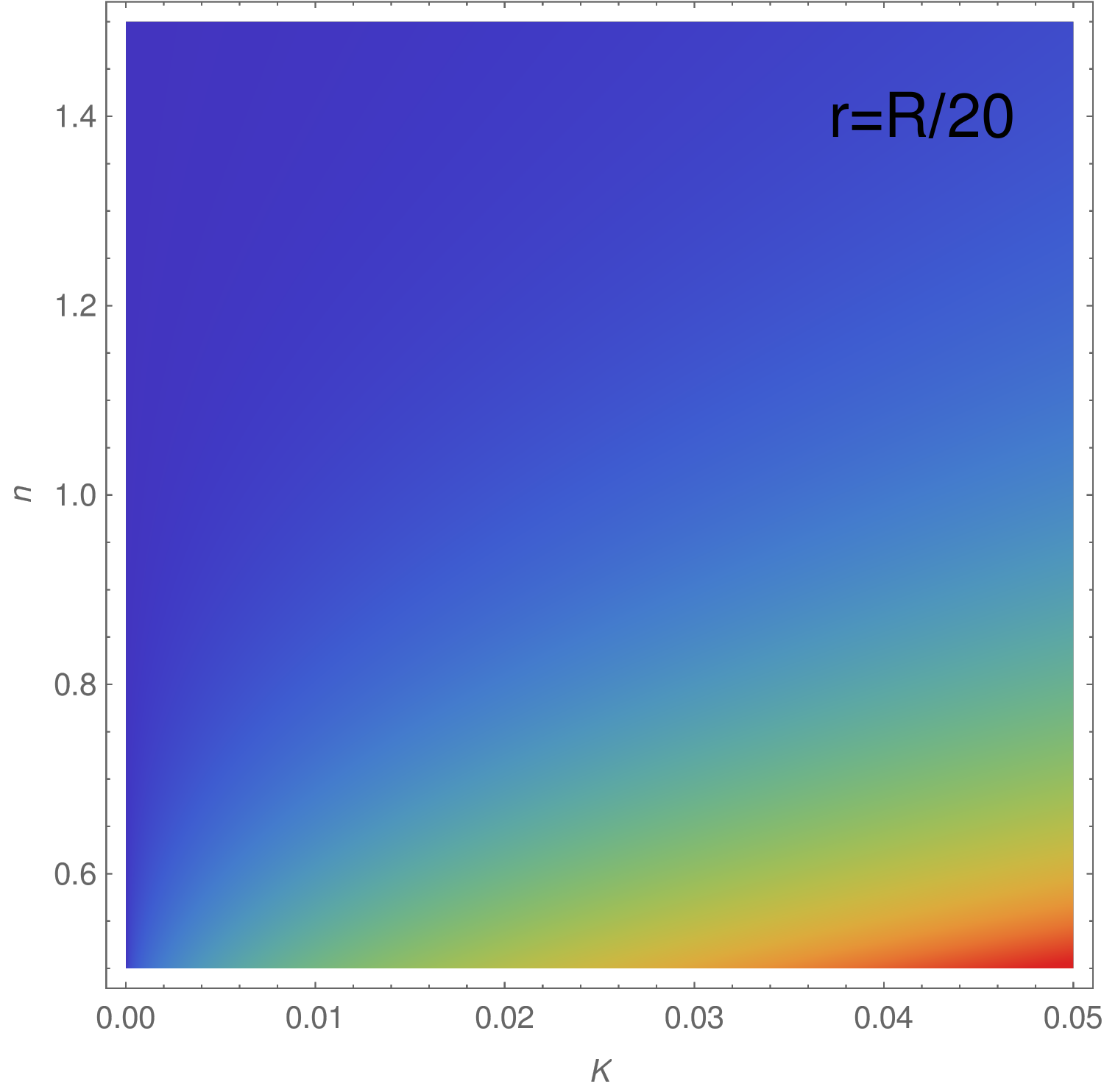} \
	\includegraphics[width=0.038\textwidth]{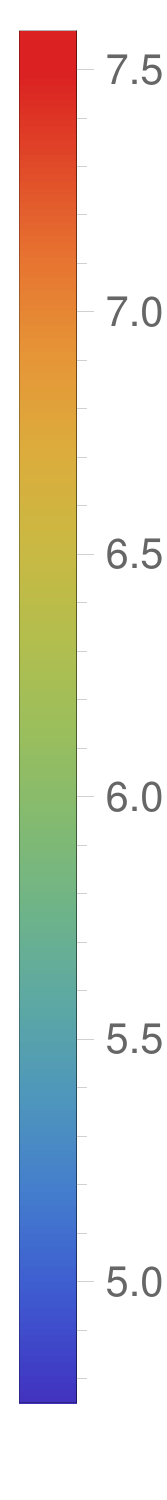} \
	\caption{Radial pressure $[\tilde{p}_r(n,K)\times\,10^4]$  for two inner layers. }
	\label{fig1}
\end{figure*}%

\item The complete process describes previously cannot be arbitrary and, in fact, is subject to the fulfillment of Bianchi identities, which implies that $\tilde{T}_{\mu\nu}=T_{\mu\nu}
+
\theta_{\mu\nu}$ is covariantly conserved, i.e., $\nabla_\mu\,\tilde{T}^{\mu\nu}
=0$. This yields the expression in Eq.~\eqref{con22}, showing an energy exchange between the relativistic fluids $\{T_{\mu\nu},\,\theta_{\mu\nu}\}$.

\end{enumerate}
We want to conclude by emphasizing two aspects that we must always keep in mind:
\begin{itemize}
	\item The GD approach is an exact scheme. 
	
	\item Regardless of the origin of $\theta_{\mu\nu}$ (as we have already mentioned, it can even represent a new gravitational sector), all our analysis is confined to the context of general relativity.
	
		\item  The sets~\eqref{ec1pf}-\eqref{ec3pf} and~\eqref{ec1d}-\eqref{ec3d} are not independent.
	
		\item The sources $\{T_{\mu\nu},\,\theta_{\mu\nu}\}$ in Einstein's equations~\eqref{ec1}-\eqref{ec3}
	can be successfully decoupled in the sets~\eqref{ec1pf}-\eqref{ec3pf} and~\eqref{ec1d}-\eqref{ec3d} if and only if the interaction between them is controlled by Eq.~\eqref{con22}.
	
	\end{itemize}

\subsection{Matching conditions at the surface}
\label{s4}
The interior ($0\le r\le R$) of the self-gravitating system of radius ($r=R$) is described by the metric~\eqref{metric}, which we can conveniently write as
\begin{equation}
	ds^{2}
	=
	e^{\nu^{-}(r)}\,dt^{2}
	-\left[1-\frac{2\,\tilde{m}(r)}{r}\right]^{-1}dr^2
	-r^{2}\,d\Omega^2
	\ ,
	\label{mgdmetric}
\end{equation}
where the interior mass function is given by
\begin{equation}
	\label{effecmass}
	\tilde{m}(r)
	=
	m(r)-\frac{r}{2}\,f(r)
	\ , 
\end{equation} 
with the Misner-Sharp mass $m$ given in Eq.~(\ref{standardGR}) and $f$ the geometric deformation in Eq.~(\ref{gd2}).
On the other hand, the exterior ($r>R$) space-time will be described by the Schwarzschild metric  
\begin{equation}
	\label{MetricSdS}
	ds^2
	=
	\left(1-\frac{2\,{\cal M}}{r}\right)\,dt^2
	-\frac{dr^2}{\left(1-\frac{2\,{\cal M}}{r}\right)}
	-r^{2}\,d\Omega^2\ . 
\end{equation}
\par
To have a smooth continuity, the metrics in Eqs.~\eqref{mgdmetric} and~\eqref{MetricSdS} must satisfy the Israel-Darmois matching
conditions at the star surface $\Sigma$ defined by $r=R$.
In particular, the continuity of the metric across $r=R$ implies
\begin{equation}
	e^{\nu ^{-}(R)}
	=
	1-\frac{2\,{\cal M}}{R}
	\ ,
	\label{ffgeneric1}
\end{equation}
and
\begin{equation}
	e^{-\lambda^{-}(R)}
	=
	1-\frac{2\,{\cal M}}{R}
	\ .
	\label{ffgeneric2}
\end{equation}

Likewise, the second fundamental form yields
\begin{equation}
	\left[\tilde{T}_{\mu \nu }\,r^{\nu }\right]_{\Sigma}
	=
	0
	\ ,
	\label{matching2x}
\end{equation}
where $r_\mu$ is the unit radial vector normal to a surface of constant $r$. Hence, using Einstein equations in Eq.~\eqref{matching2x}, we have
\begin{equation}
	\left[p_r+{\cal P}_r\right]_{\Sigma }
	=
	0
	\ .
	\label{matching3}
\end{equation}
This matching condition takes the final form 
\begin{equation}
	p_{R}
	+{\cal P}_R
	=
	0
	\ ,
	\label{matchingf}
\end{equation}
where $p_{R}\equiv p(R)$ and ${\cal P}_R\equiv\,{\cal P}(R)\,$.
The condition~\eqref{matchingf} can be written as
\begin{equation}
	\tilde{p}_R\equiv\,p_{R}+\frac{f_{R}}{\kappa\,}
	\left(\frac{1}{R^{2}}+\frac{\nu _{R}^{\prime }}{R}\right)+\frac{g_{R}^{\prime }}{\kappa\,\,R}e^{-\mu}=0
	\ ,
	\label{sfgeneric}
\end{equation}
where $\nu _{R}^{\prime }\equiv \partial _{r}\nu^{-}|_{r=R}$. Eqs.~(\ref{ffgeneric1}), (\ref{ffgeneric2})
and~(\ref{sfgeneric}) are the necessary and sufficient conditions for matching the interior
GD metric~(\ref{mgdmetric}) with the outer Schwarzschild metric~\eqref{MetricSdS}.
\par
\section{Polytropic equation of state}
\label{sec3}
\begin{figure*}[ht!]
	\centering
	\includegraphics[width=0.28\textwidth]{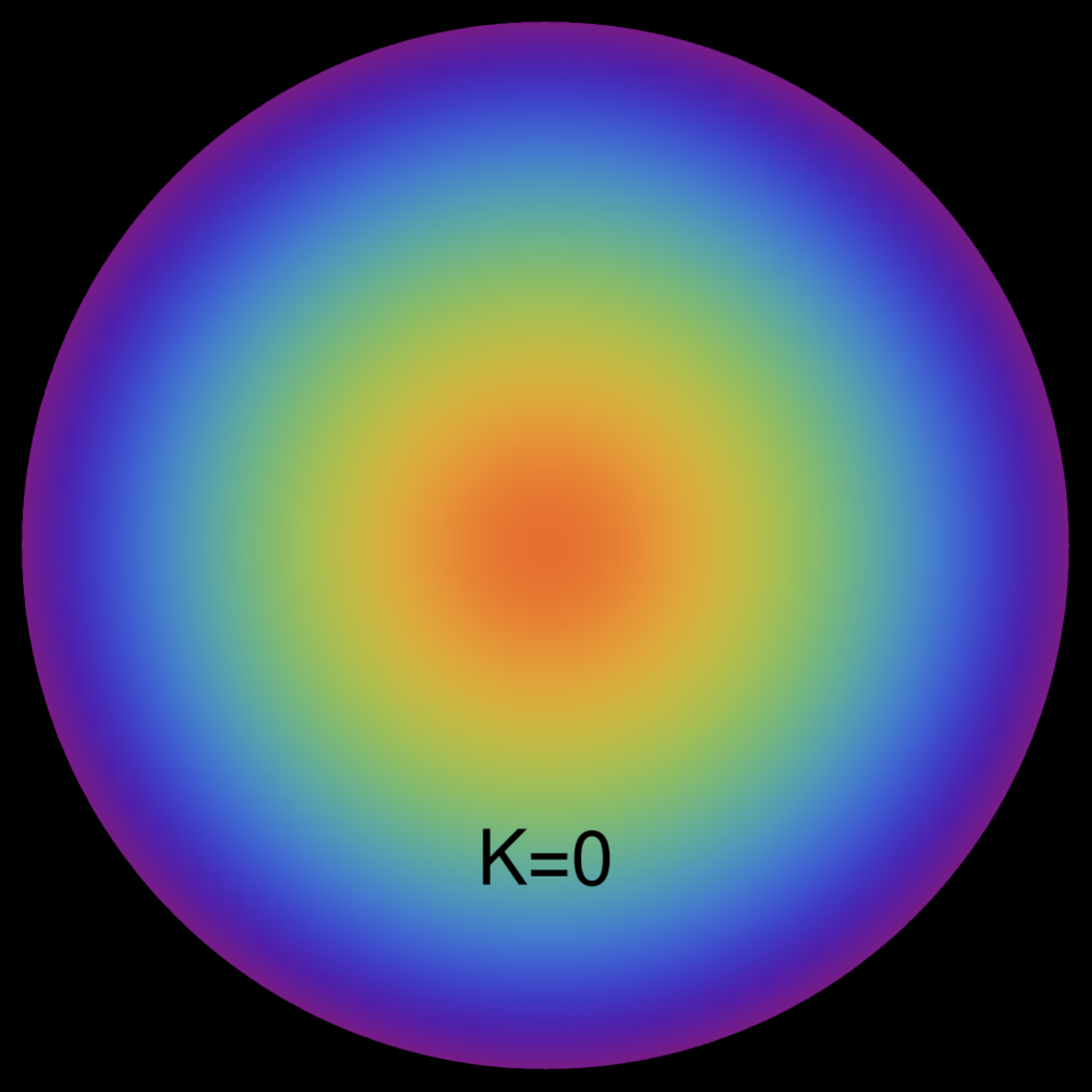} \
	\includegraphics[width=0.025\textwidth]{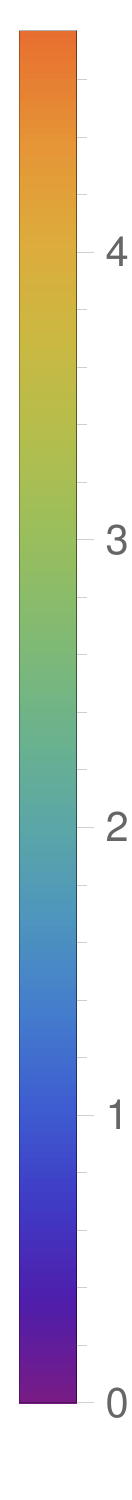} \
	\includegraphics[width=0.28\textwidth]{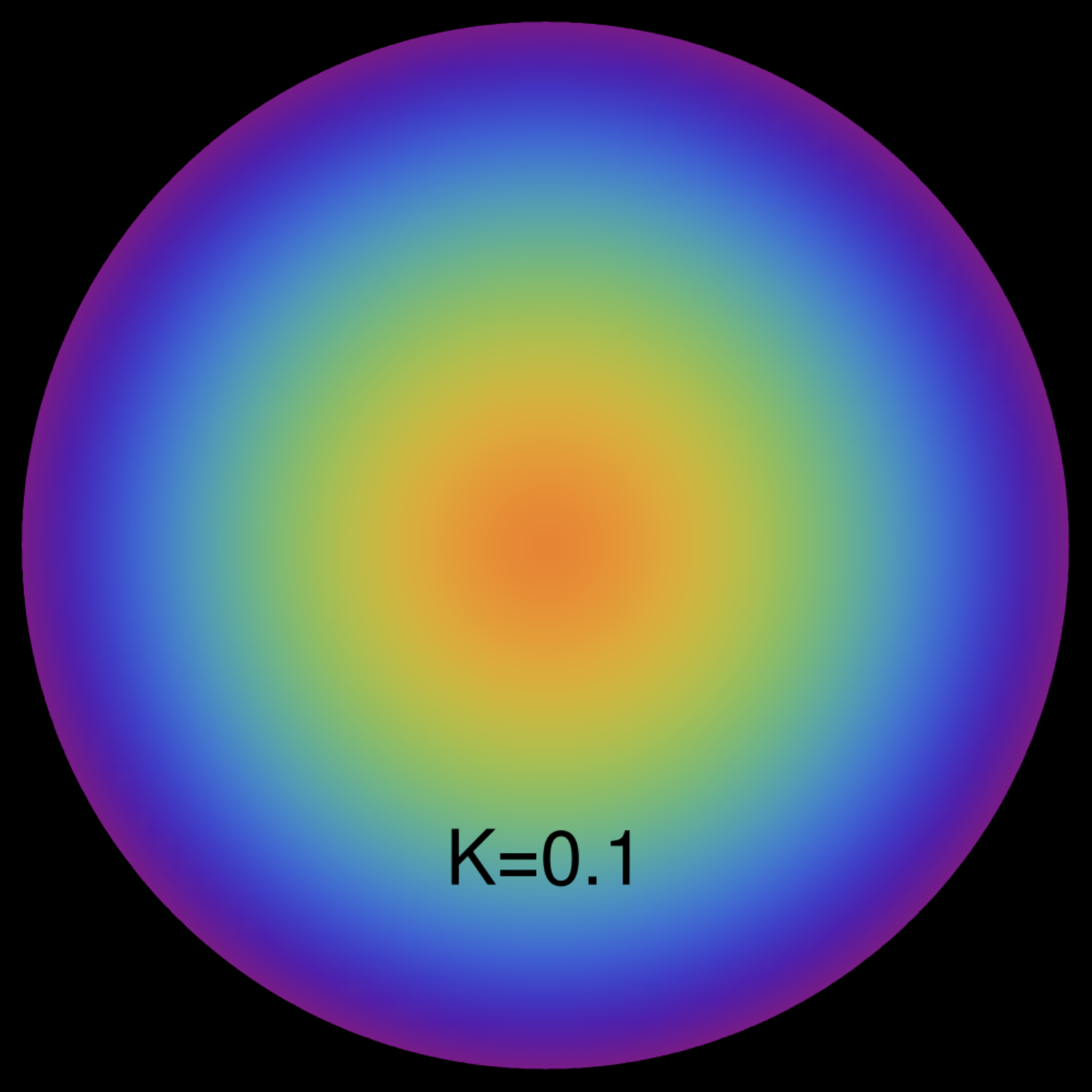} \
	\includegraphics[width=0.025\textwidth]{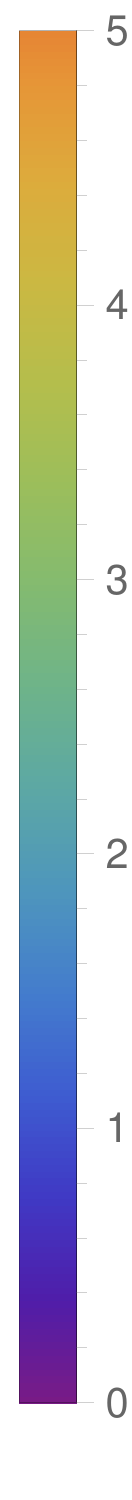} \
	\includegraphics[width=0.28\textwidth]{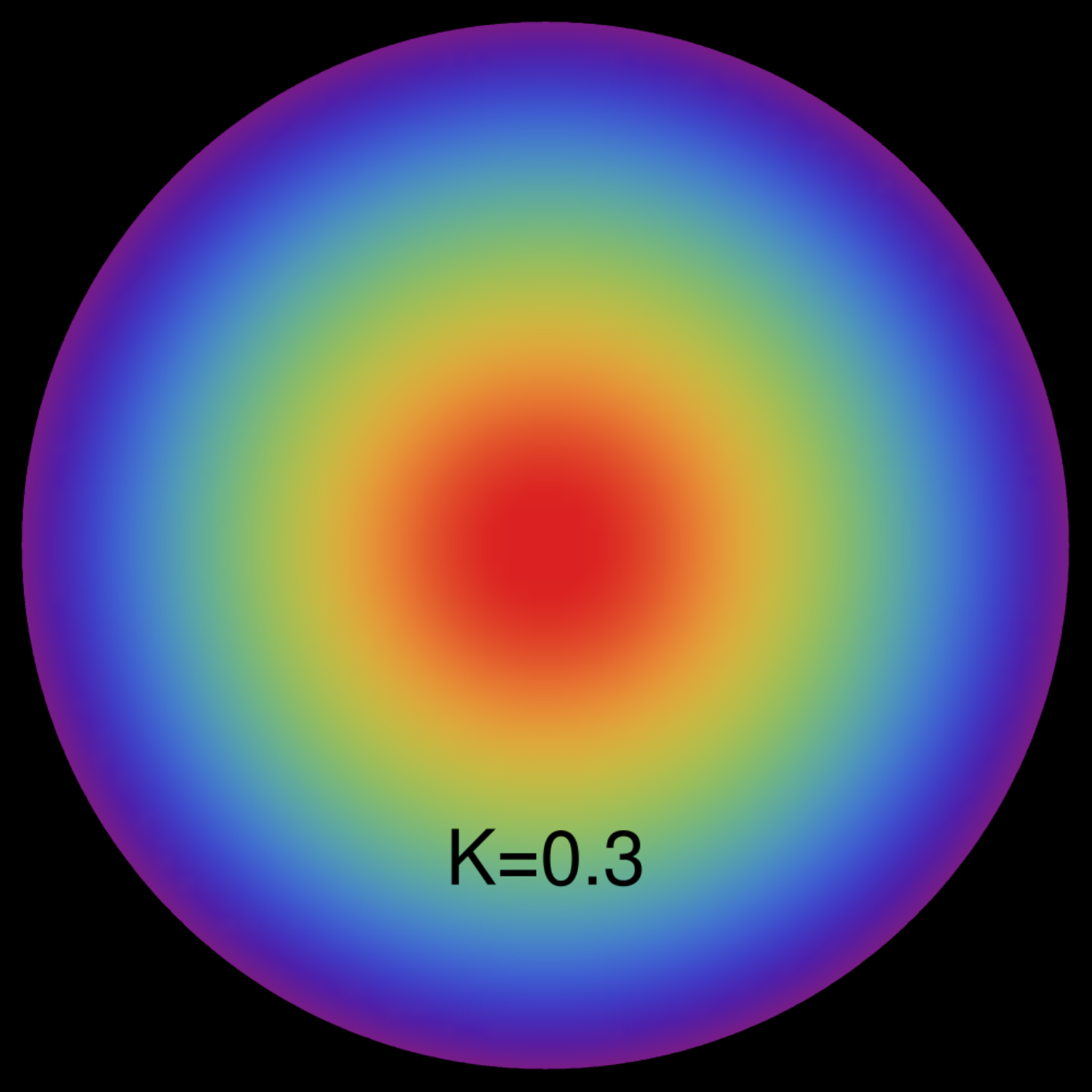} \
	\includegraphics[width=0.025\textwidth]{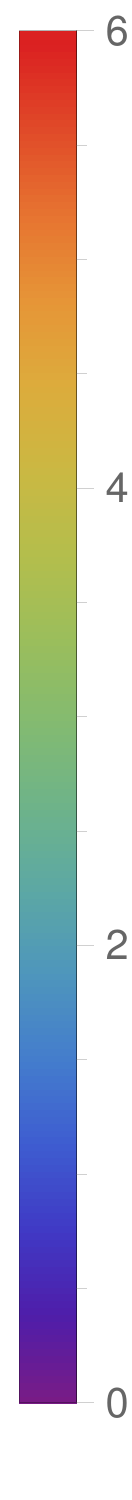} \
	\caption{Radial pressure $[\tilde{p}_r(r)\times\,10^4]$ for three different cases, where we can see the effects of the polytrope on the  perfect fluid ($K=0$). We take $n=3$}
	\label{fig2}
\end{figure*}%
So far, all our analysis has been generic, without specifying the sources $\{T_{\mu\nu},\,\theta_{\mu\nu}\}$ that compose our system. Of all the possible gravitational sources, we will choose one of particular importance, which has been extensively investigated. We refer to a polytropic fluid, which in our case will be represented by the tensor $\theta_{\mu\nu}$. Hence, following our previous analysis, we will see how to elucidate the effects of a polytrope on another generic source $T_{\mu\nu}$ {describe by Einstein's equations~\eqref{ec1pf}-\eqref{ec3pf}.
\par
If the tensor $\theta_{\mu\nu}$ represents an isotropic polytrope, it satisfies the equation of state
\begin{equation}
	\label{o}
	{\cal P}_r={\cal P}_t
	=
	K
	\left({\cal E}\right)^\Gamma
	\ .
\end{equation}
However, in our case we will require that only the radial pressure satisfies the equation of state~\eqref{o}, allowing the tangential component to evolve independently. Hence,
\begin{equation}
	\label{polyt0}
	{\cal P}_r
	=
	K
	\left({\cal E}\right)^\Gamma\neq\,{\cal P}_t
	\ ,
\end{equation}
with $\Gamma=1+1/n$, where $n$ is the polytropic index and $K>0$ denotes a parameter which
contains the temperature implicitly and is governed by the thermal characteristics of a given polytrope. (For all details regarding basic concepts of polytropes, see for instance Ref.~\cite{Horedt:2004pol}, also see references Refs.~\cite{Herrera:2013fja,Stuchlik:2016xiq,Stuchlik:2000gey,Novotny:2017cep,Stuchlik:2017qiz,Posada:2020svn}).). 
\par
Let us start by using Eqs.~(\ref{ec1d}) and (\ref{ec2d}) in the expression~(\ref{polyt0}), which yields a first order
non-linear differential equation for the deformation $f$,  
\begin{equation}
	\label{poly2}
	\frac{{f}'}{r}
	+
	\frac{f}{r^2}
	=-\left(\frac{\kappa^{\Gamma-1}}{K}\right)^{1/\Gamma}\left[\frac{e^{-\mu}\,g'}{r}+f\left(\frac{1}{r^2}+\frac{\nu'}{r}\right)\right]^{1/\Gamma}
\end{equation}
 Therefore, given a seed solution $\{\xi,\,\mu\}$ to Einstein equations~\eqref{ec1pf}-\eqref{ec3pf}, we end with a non-linear differential expression in Eq.~\eqref{poly2} to determinate the deformations $\{g,\,f\}$. Hence, we need to prescribe additional information. In any case, we must be careful in keeping the physical acceptability of the seed solution $\{\xi,\,\mu\}$, which is not a trivial issue. In this respect, and in order to ensure the coupling condition in Eq.~\eqref{sfgeneric}, we impose the so-called mimic constraint for the pressure, namely,
\begin{equation}
	\label{mimic}
{\cal P}_r\sim\,{p}_{r}
	\ .
\end{equation}
The simplest expression for ${\cal P}_r(r)$ satisfying the constraint~\eqref{mimic} is given by
\begin{equation}
	\label{mimic2}
	{\cal P}_r(r)=\alpha(K,\,\Gamma){p}_{r}(r)\ ,
\end{equation}
where $\alpha(K,\,\Gamma)$ is a characteristic function for each polytrope. The simplest form for $\alpha(K,\,\Gamma)$ consistent with the polytropic equation of state~\eqref{polyt0} and with the condition
\begin{equation}
	\label{jr}
	f(r)\big|_{K=0}=0
\end{equation}
is given by
\begin{equation}
	\label{mimic3}
\alpha(K,\,\Gamma)=K^{\Gamma}\ .
\end{equation}
Hence, the expression~\eqref{mimic2} becomes 
\begin{equation}
	\label{mimic4}
	{\cal P}_r(r)=K^{\Gamma}\,{p}_{r}(r)\ ,
\end{equation}
Expressions in Eqs.~\eqref{poly2} and~\eqref{mimic4} now are written as
\begin{eqnarray}
	\label{poly3}
	&&\frac{{f}'}{r}
	+
	\frac{f}{r^2}
	=-\,(\kappa\,K)^{\frac{\Gamma-1}{\Gamma}}\left[e^{-\mu}\left(\frac{1}{r^2}+\frac{\xi'}{r}\right)-\frac{1}{r^2}\right]^{1/\Gamma}\ ,
	\\
	\label{poly4}
	&&g'=\left(\frac{1}{e^{-\mu}+f}\right)\left[\left(K^{\Gamma}\,e^{-\mu}-f\right)\left(\xi'+\frac{1}{r}\right)-\frac{K^{\Gamma}}{r}\right]\ .
	\nonumber\\
\end{eqnarray}
We see that for a given seed solution $\{\xi,\,\mu\}$ to Einstein equations~\eqref{ec1pf}-\eqref{ec3pf},  we can determinate its deformation $\{g,\,f\}$ produced for any polytrope $\{K,\,\Gamma\}$ by Eqs.~\eqref{poly3} and~\eqref{poly4}. 

\begin{figure*}[ht!]
	\centering
	\includegraphics[width=0.28\textwidth]{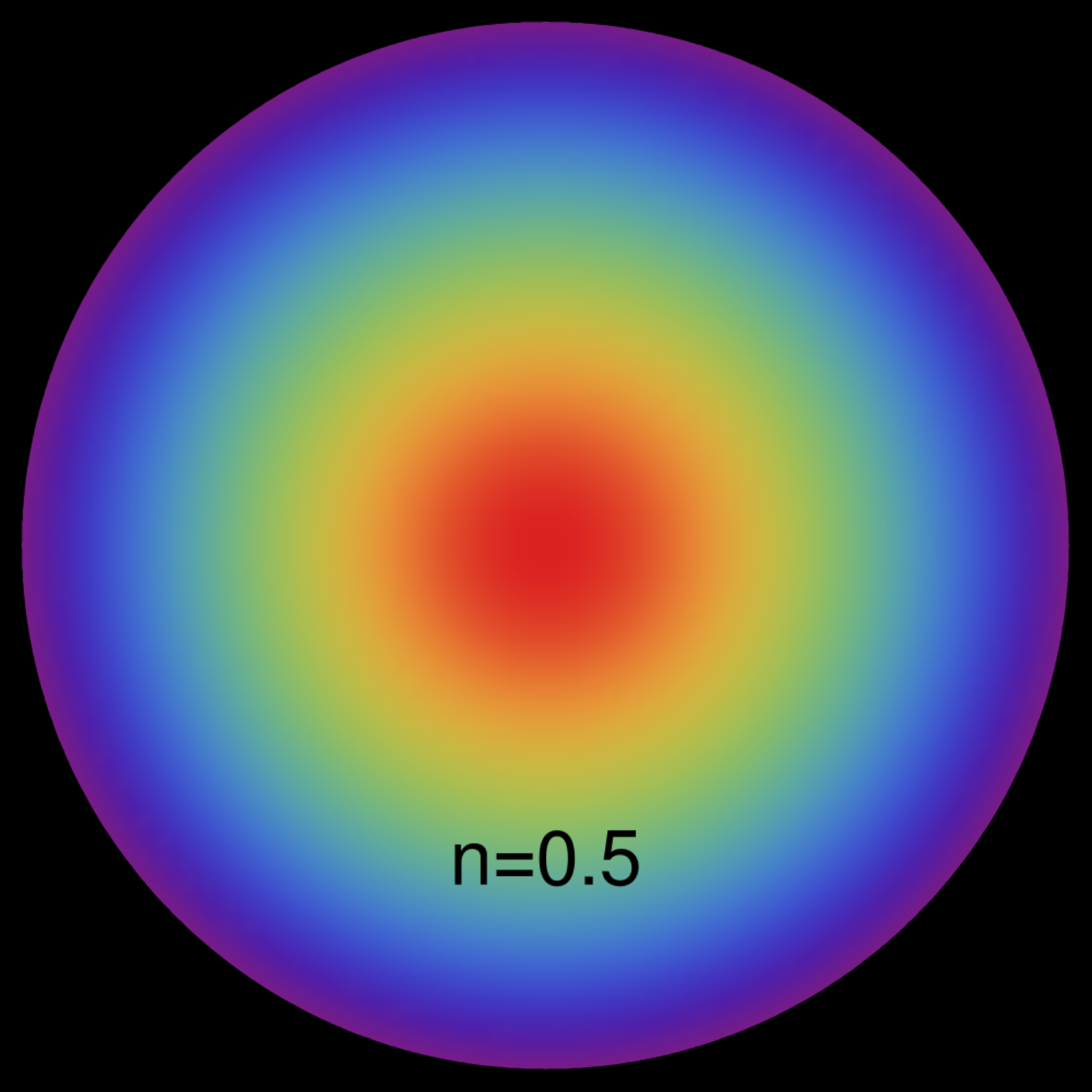} \
	\includegraphics[width=0.025\textwidth]{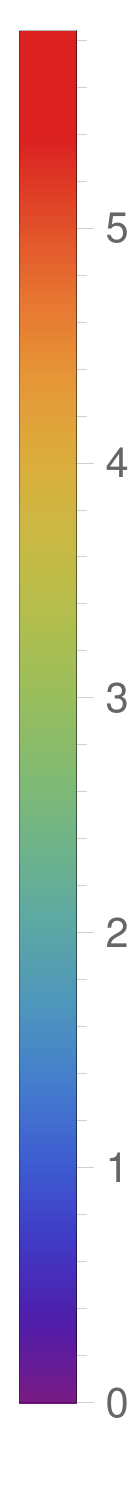} \
	\includegraphics[width=0.28\textwidth]{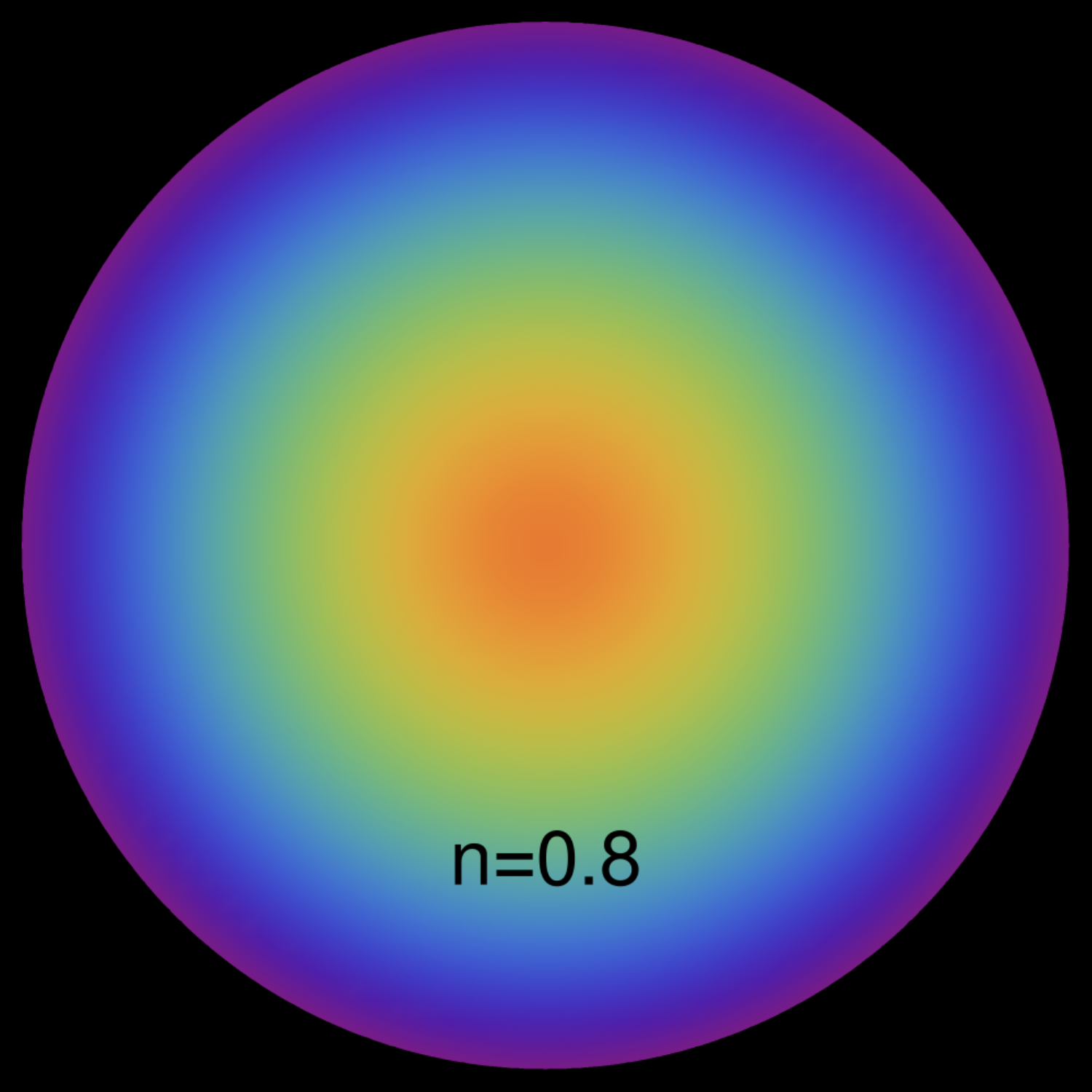} \
	\includegraphics[width=0.025\textwidth]{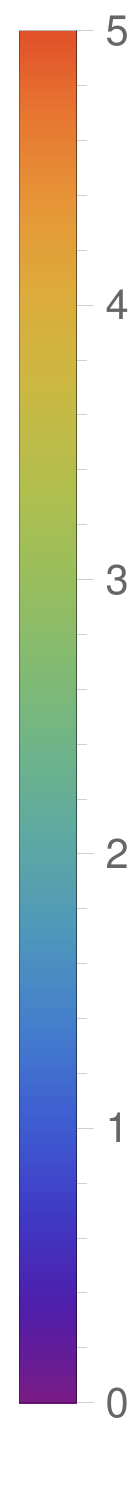} \
	\includegraphics[width=0.28\textwidth]{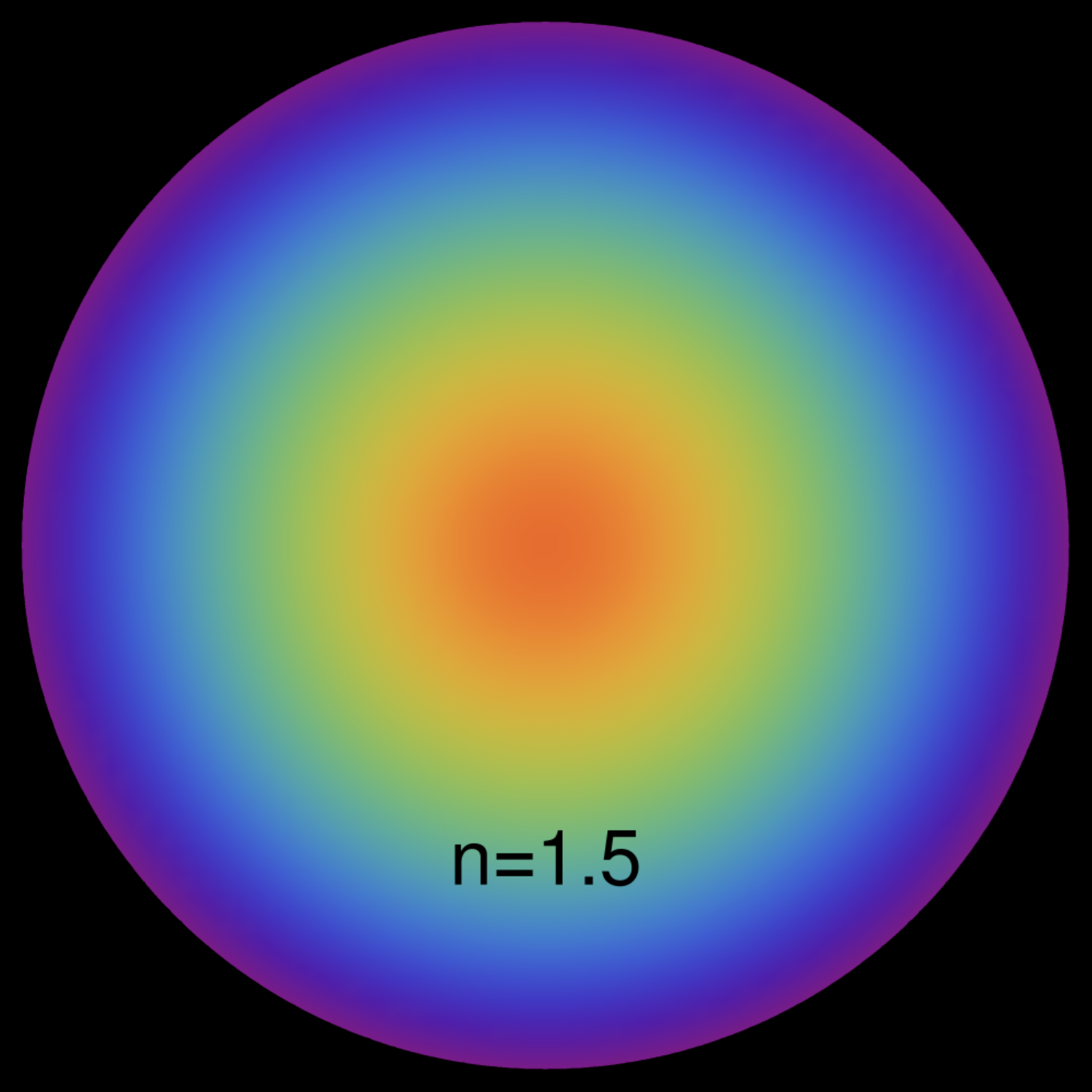} \
	\includegraphics[width=0.025\textwidth]{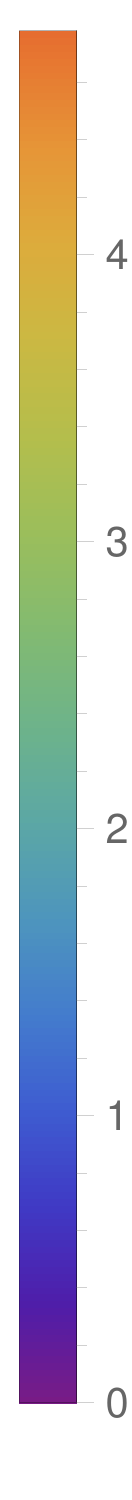} \
	\caption{Radial pressure $[\tilde{p}_r(r)\times\,10^4]$ for $K=0.01$}
	\label{fig3}
\end{figure*}%

A condition other than~\eqref{mimic}, also useful to ensure a physically acceptable solution, is to impose the so-called mimic constraint for the pressure, i.e,
\begin{equation}
	\label{mimicd}
	{\cal E}\sim\,\rho
	\ .
\end{equation}
Hence, following the same reasoning as Eqs.~\eqref{mimic2}-\eqref{mimic3}, we have
\begin{equation}
	\label{mimicd2}
	{\cal E}(r)=K^{\Gamma}\,\rho(r)\ ,
\end{equation}
which yields, 

\begin{eqnarray}
	\label{poly3d}
	&&\frac{{f}'}{r}
	+
	\frac{f}{r^2}
	=-\alpha\left[\frac 1{r^2}
	-
	e^{-\mu }\left( \frac1{r^2}-\frac{\mu'}r\right)\right]\ ,
	\\
	\label{poly4d}
	&&g'=\left(\frac{r}{e^{-\mu}+f}\right)\left[\kappa\,K\left(\frac{\alpha}{\kappa}\left[\frac 1{r^2}
	-
	e^{-\mu }\left( \frac1{r^2}-\frac{\mu'}r\right)\right]\right)^\Gamma\right.
	\nonumber\\	
	&&\left. -f\left(\frac{1}{r^2}+\frac{\xi'}{r}\right)\right]\ .
	\nonumber\\
\end{eqnarray}
 In short, our approach allows to determinate the effects of politropes on any generic fluid, represented by $T_{\mu\nu}$ and satisfying Einstein equations~\eqref{ec1pf}-\eqref{ec3pf}, no matter its nature.
\par
Note that if we impose the constraint~\eqref{mimic2}, we are faced with solving the nonlinear differential equation~\eqref{poly3} to determine $f(r)$. Instead, if we impose the condition~\eqref{mimicd2}, we will need to solve the linear differential equation~\eqref{poly3d} to find $f(r)$. Everything seems to indicate that it is more convenient to impose the constraint~\eqref{mimicd2}. However, using the condition~\eqref{mimic2} has a quite useful advantage: the coupling problem on the surface, which could be non-trivial in some cases, is greatly reduced. 

Finally, we see that a critical characteristic of the interaction between both fluids, such as the exchange of energy-momentum  $\Delta\,{\rm E}$  between them, is easily elucidated by [see Eq.~\eqref{con22}]
\begin{eqnarray}
	\label{exchange}
\Delta\,{\rm E}=
	\frac{g'}{2}\left(\rho+p_r\right)
	\ ,
\end{eqnarray}
which we can write in terms of pure geometric functions as [see Eqs.~\eqref{ec1pf}-\eqref{ec3pf}]
\begin{eqnarray}
	\label{exchange2}
	\Delta\,{\rm E}=
	\frac{g'}{2\,\kappa}\frac{e^{-\mu}}{r}\left(\xi'+\mu'\right)
	\ .
\end{eqnarray}
From the expression~\eqref{exchange} we can see that $g'>0$ yields $\Delta\,{\rm E}>0$. This indicates $\nabla_\sigma\theta^{\sigma}_{\ \nu}>0$,  according to the conservation equation~\eqref{con22}, which means that the polytrope is giving energy to the environment. The opposite happens when $g'<0$.
\begin{figure*}[ht!]
	\centering
	\includegraphics[width=0.29\textwidth]{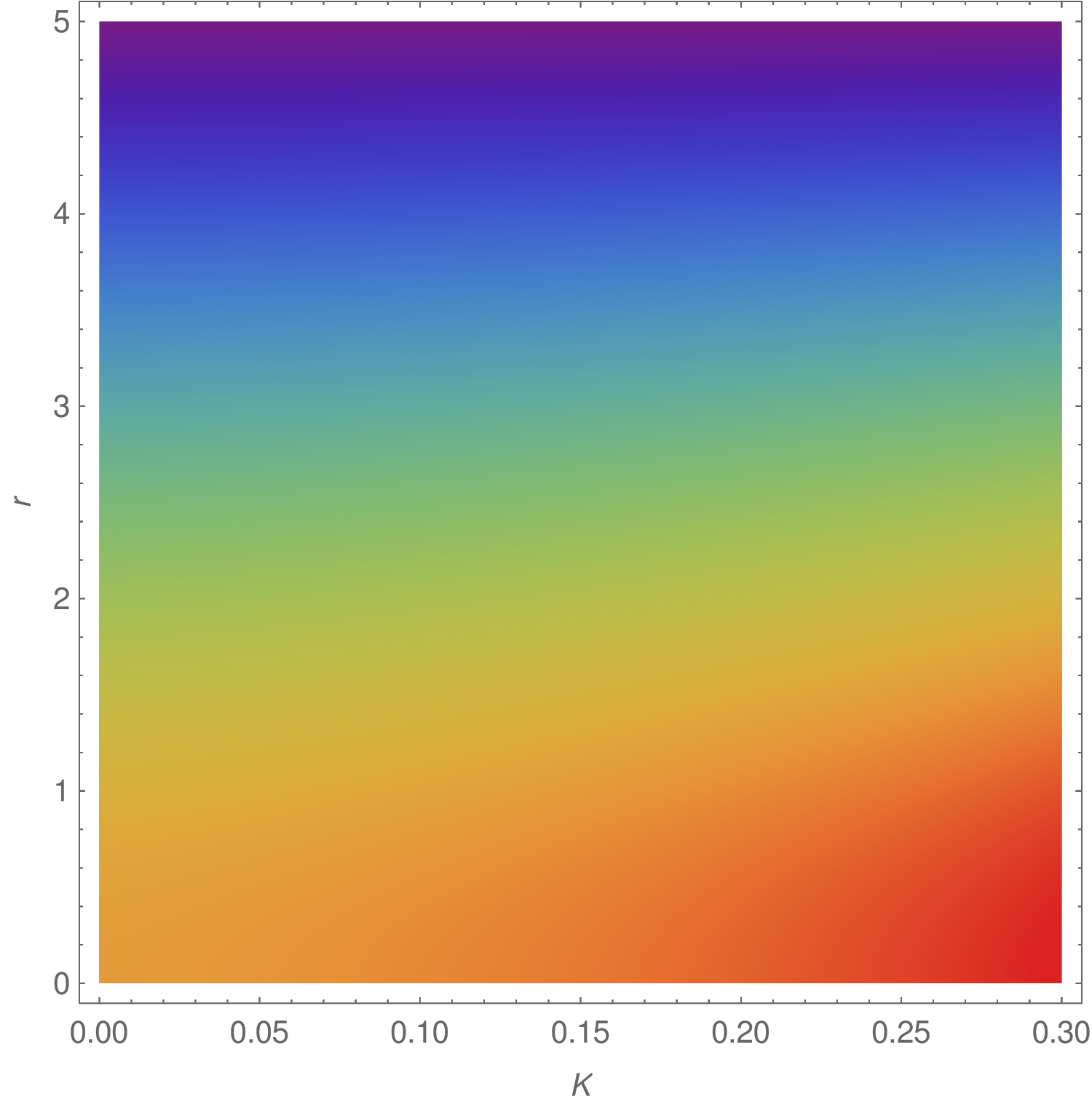} \
	\includegraphics[width=0.026\textwidth]{plot8a} \
	\includegraphics[width=0.29\textwidth]{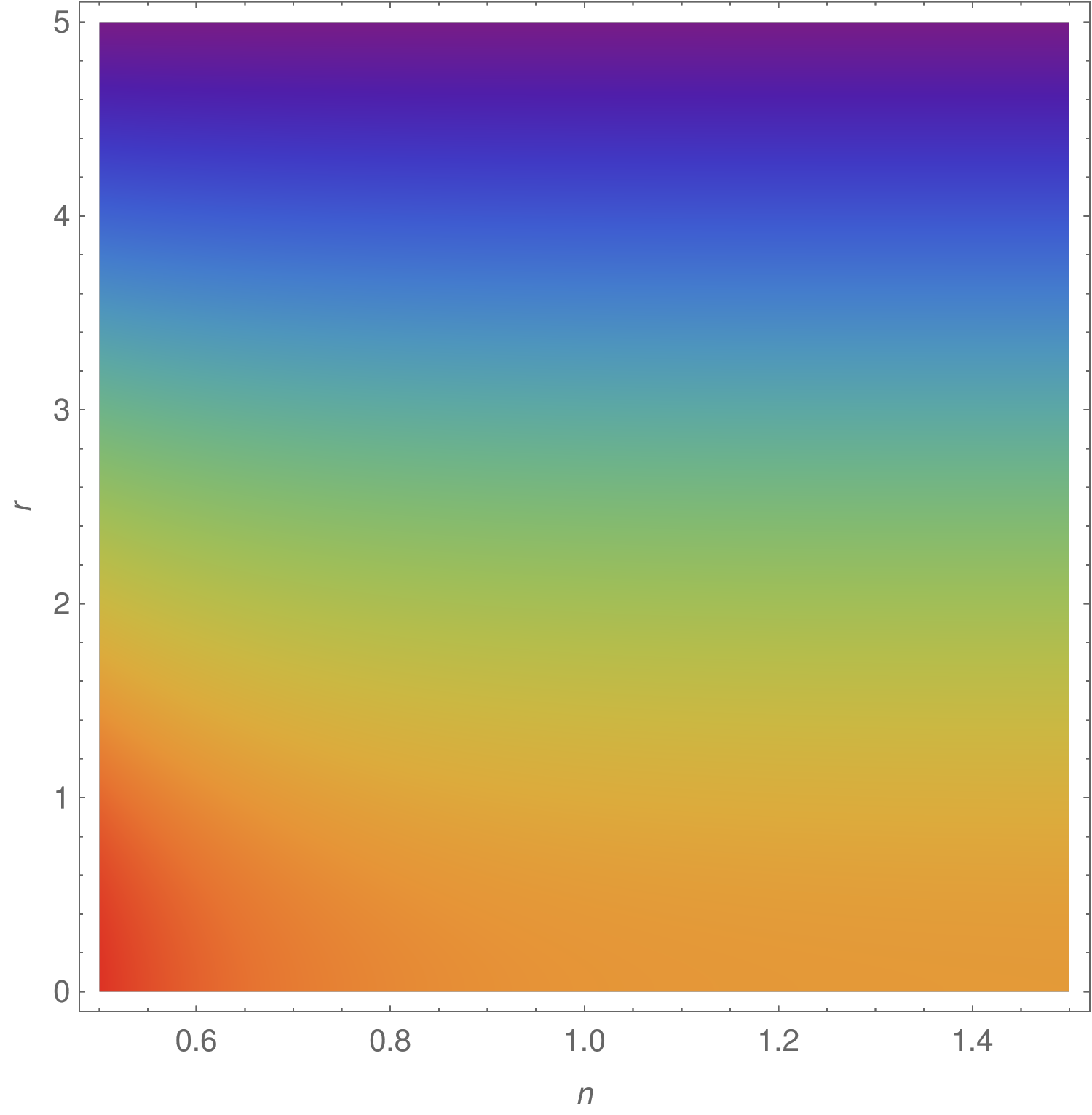} \
	\includegraphics[width=0.026\textwidth]{plot3a} \
	\caption{Radial pressure $[\tilde{p}_r(r,K)\times\,10^4]$ (left panel) and $[\tilde{p}_r(r,n)\times\,10^4]$ (right panel).}
	\label{figx}
\end{figure*}%
\subsubsection{Strategy}
\label{protocol}
We can now detail our scheme to elucidate the effects of the polytrope $\{{\cal E},\,{\cal P}_r,\,{\cal P}_t\}$ on any other generic fluid $\{{\rho},\,{p}_r,\,{p}_t\}$, no matter its nature (isotropic, charged, scalar field, etc.):
\begin{enumerate}
	\item
	Take any solution~\eqref{pfmetric} of the Einstein field equations ${G}_{\mu\nu}=\kappa\,{T}_{\mu\nu}$ in Eqs.~\eqref{ec1pf}-\eqref{ec3pf}.
	\item Consider a polytropic fluid, characterized by the constant $K$ and index $n$ in the equation of state~\eqref{polyt0}.
	\item
Impose the condition~\eqref{mimic4} [or Eq.~\eqref{mimicd2}] to ensure a polytropic fluid with acceptable physical behavior. 
	\item Use the metric components $\{\xi,\,\mu\}$ displayed in Eq.~\eqref{pfmetric} to find $\{f,\,g'\}$ by Eqs.~\eqref{poly3} and~\eqref{poly4} [or by Eqs.~\eqref{poly3d} and~\eqref{poly4d} ] (it is NOT necessary to find $g$).
	\item Now use $\{f,\,g'\}$ in the field equations~\eqref{ec1d}-\eqref{ec3d} to explicitly determinate the polytrope $\{{\cal E},\,{\cal P}_r,\,{\cal P}_t\}$.  
	\item Finally, use the metric~\eqref{pfmetric} and $\{f,\,g'\}$ to find $\{\tilde{\rho},\,\tilde{p}_r,\,\tilde{p}_t\}$ through Einstein equations~\eqref{ec1}-\eqref{ec3}. Hence, at this stage, we can elucidate the effects of polytropes $\{{\cal E},\,{\cal P}_r,\,{\cal P}_t\}$ on any other source $\{{\rho},\,{p}_r,\,{p}_t\}$ by simple inspection of the total effective fluid  $\{\tilde{\rho},\,\tilde{p}_r,\,\tilde{p}_t\}$ [and its space-time, described by the metric~\eqref{metric}].
\end{enumerate}
We want to emphasize that the previous scheme allows us to study the coexistence of a polytropic fluid with any other, and elucidate the effects of the former on the latter, by a systematic and direct way. Next we will consider a perfect fluid as a seed solution to elucidate the consequences of polytrope on this gravitational sources. 

\section{Coexistence of polytropes\\ and perfect fluids}
\label{sec4}

In particular, we can simply choose a known solution with physical relevance, like the well-known Tolman~IV solution
$\{\xi,\mu,\rho, p\}$ for perfect fluids~\cite{Tolman:1939jz}, namely, 
\begin{eqnarray}
	\label{tolman00}
	&&e^{\xi(r)}
	=
	B^2\,\left(1+\frac{r^2}{A^2}\right)
	\ ,
	\\
	\label{tolman11}
	&&e^{-\mu(r)}
	=
	\frac{\left(1-\frac{r^2}{C^2}\right)\left(1+\frac{r^2}{A^2}\right)}{1+\frac{2\,r^2}{A^2}}
	\ ,
	\\
	\label{tolmandensity}
	&&\rho(r)
	=
	\frac{3A^4+A^2\left(3C^2+7r^2\right)+2 r^2 \left(C^2+3 r^2\right)}{\kappa\,C^2\left(A^2+2r^2\right)^2}
	\ ,
	\\
	&&p(r)=
	\frac{C^2-A^2-3r^2}{\kappa\,C^2\left(A^2+2r^2\right)}
	\label{tolmanpressure}
	\ .
\end{eqnarray}
The constants $A$, $B$ and $C$ in Eqs.~(\ref{tolman00})-(\ref{tolmanpressure}) are determined by the matching conditions in Eqs.~(\ref{ffgeneric1}), (\ref{ffgeneric2}) 
and~(\ref{sfgeneric}) [with $f_R=g_R=0$] between the above interior solution and
the exterior metric in Eq.~\eqref{MetricSdS}.
This yields
\begin{equation}
	\label{A}
	\frac{A^2}{R^2}
	=
	\frac{R-3\,M}{M}
	\ ,
	\qquad
	B^2=1-\frac{3\,M}{R}
	\ ,
	\qquad
	\frac{C^2}{R^2}
	=
	\frac{R}{M}
	\ ,
\end{equation}
with the compactness $M/R<4/9$, and $M=m(R)$ the total mass in Eq.~(\ref{standardGR}).
The expressions in Eq.~(\ref{A}) ensure the geometric continuity at $r=R$ and will change when we add the polytrope source $\theta_{\mu\nu}$ [indeed, the constant A in Eq.~\eqref{A} will change as $A\to\,A(K,n)$].
\begin{figure*}[ht!]
	\centering
	\includegraphics[width=0.33\textwidth]{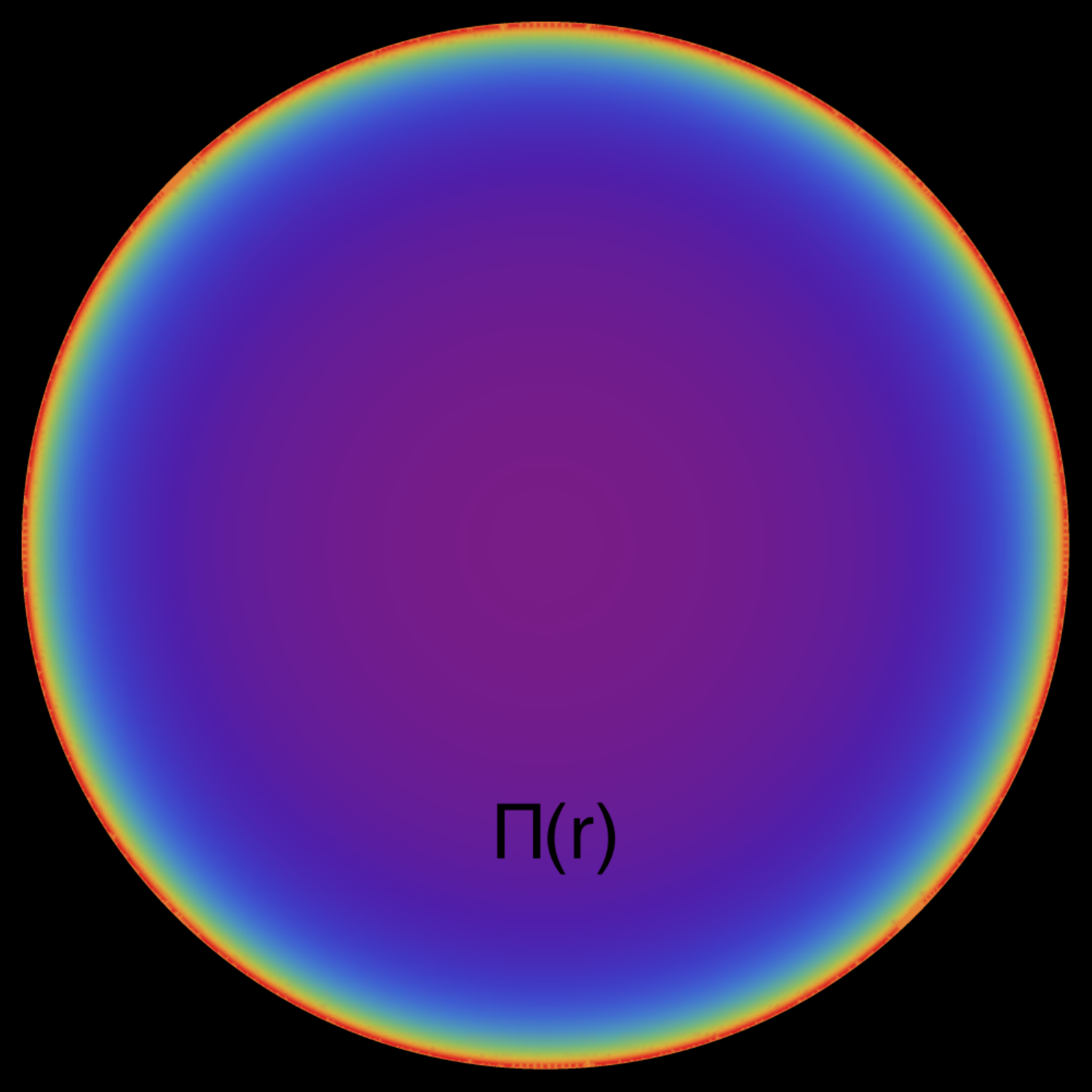} \
	\includegraphics[width=0.0295\textwidth]{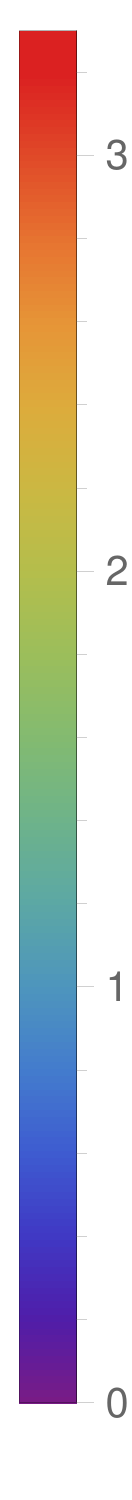} \
	\includegraphics[width=0.33\textwidth]{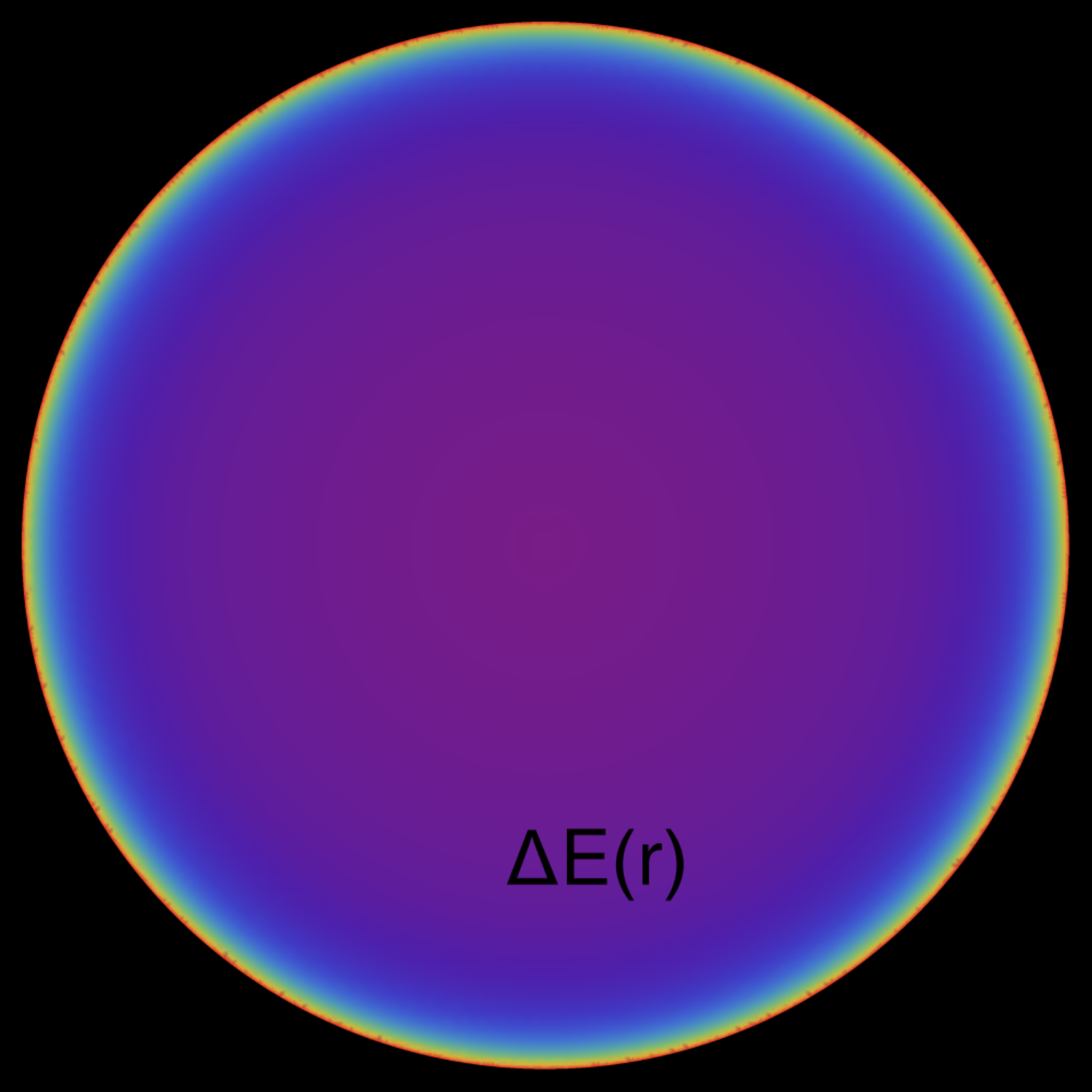} \
	\includegraphics[width=0.037\textwidth]{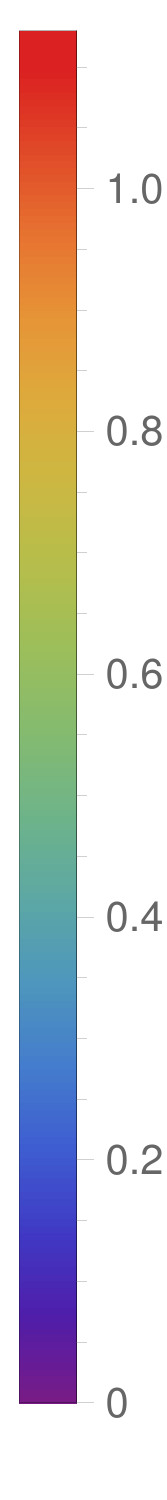} \
	\caption{Anisotropy $[\Pi(r)\times\,10^2]$ and exchange of energy $[\Delta{\rm E}(r)\times\,10^2]$  for $n=0.5$ and $K=0.01$}
	\label{fig4}
\end{figure*}%
Using the metric functions in Eqs.~\eqref{tolman00} and~\eqref{tolman11} in the differential expression~\eqref{poly3} we obtain the geometric deformation in terms of the polytropic index $n$, which reads
\begin{equation}
	\label{deforp}
	f(r)
	=-\frac{1}{3}(\kappa\,K)^{\frac{1}{n+1}}r^2\left(\frac{C^2-A^2}{A^2\,C^2}\right)^{\frac{n}{n+1}}\,F_n(r)+\frac{c_1}{r}
	\ ,
\end{equation}
where $F_n(r)$ is an Appell hypergeometric function and the integration constant $c_1=0$ to have a regular solution in the origin $r=0$. Let us remind that, contrary to the radial metric component $\lambda$, the temporal one $\nu$ appears only as functions of its derivatives in Einstein equations~\eqref{ec1}-\eqref{ec3}. In this sense, to determine the source of the metric~\eqref{metric}, it is not necessary to obtain the explicit form of the temporal deformation $g$ by Eq.~\eqref{poly4}. 

The continuity of the first fundamental form given by Eqs.~\eqref{ffgeneric1} and~\eqref{ffgeneric2} leads to
\begin{equation}
	B^2\,\left(1+\frac{R^2}{A^2}\right)e^{g_R}
	=
	1-\frac{2\,{\cal M}}{R}
	\ ,
	\label{ffgeneric1p}
\end{equation}
and
\begin{equation}
	e^{-\mu}+f_{R}
	=
	1-\frac{2\,{\cal M}}{R}
	\ ,
	\label{ffgeneric2p}
\end{equation}
where $f_{R}=f(R)$ is the deformation evaluated at the star surface. The continuity of the second fundamental form in Eq.~\eqref{sfgeneric} yields
\begin{equation}
	\label{C}
	C^2=A^2+3R^2
\end{equation}
and then the deformation in Eq.~\eqref{deforp} takes the final form
\begin{equation}
	\label{deffin}
	f(r)
	=-\left(\frac{\kappa\,K}{3}\right)^{\frac{1}{n+1}}r^2\left[\frac{R^2}{A^2(A^2+3R^2)}\right]^{\frac{n}{n+1}}\,F_n(r)
	\ ,
\end{equation}
On the other hand, by using the condition in~\eqref{ffgeneric2p}, we obtain for the Schwarzschild mass
\begin{equation}
	\label{Smass}
\frac{2\,{\cal M}}{R}=\frac{2\,M}{R}-f_{R}
	\ ,
\end{equation}
where $M=m(R)$ in the expression in Eq.~\eqref{standardGR} has been used. Finally, by using the expression in Eq.~\eqref{Smass} in the matching condition~\eqref{ffgeneric1p}, we obtain
\begin{equation}
	B^2\,\left(1+\frac{R^2}{A^2}\right)e^{g_R}
	=\frac{A^2+R^2}{A^2+3R^2}+f_{R}
	\ .
	\label{ffgeneric1pp}
\end{equation}
Eqs.~\eqref{C},~\eqref{Smass} and~\eqref{ffgeneric1pp} are the necessary and sufficient conditions for the matching of the
interior metric~\eqref{metric} to a spherically symmetric outer “vacuum” described by the Schwarzschild metric in Eq.~\eqref{MetricSdS}. From equation~\eqref{ffgeneric1pp} we see that the constants in Eq.~\eqref{A} are now functions of the polytropic variables, that is,
\begin{equation}
	A\rightarrow\,A(K,\Gamma)\ ,\,\,\,\,	B\rightarrow\,B(K,\Gamma)\ ,\,\,\,\,	C\rightarrow\,C(K,\Gamma)\ .
\end{equation}
Also notice that for a given polytrope $\{K,\,n\}$, the expression in Eq.~\eqref{ffgeneric1pp} contains two unknown functions $\{A,\,B\}$. We might be tempted to eliminate $B$ by a time rescaling $t\to\,\tilde{t}=Bt$ in the metric~\eqref{metric}, but this would lead to a solution where the perfect fluid in Eqs.~\eqref{tolman00}-\eqref{A} is not regained when $g=f=0$. Since we want to keep the Tolman IV solution in this limit, we introduce
\begin{equation}
	\label{AA}
	A(K,\Gamma)=A_0+\zeta(K,\Gamma)\ ,
\end{equation}
where $A_0$ is the perfect fluid value in Eq.~\eqref{A}, and $\zeta(K,\Gamma)$ a function with
dimensions of a length encoding the polytropic effects, which satisfies
\begin{equation}
	\label{klimit}
	\zeta(K,\Gamma)\big\vert_{K=0}=0\ .
\end{equation}
Hence, given an expression for $\zeta(K,\Gamma)$, we can determinate $B(K,\Gamma)$ by the condition~\eqref{ffgeneric1pp}, so that the problem at the stellar surface is closed. We want to conclude by emphasizing that the expression in Eq.~\eqref{AA} does not mean any approximation, much less a perturbative analysis.

Next we will proceed with a simple reasonable expression for $\zeta$, given by
\begin{equation}
	\label{zeta}
	\zeta(K,\Gamma)=-\frac{R}{M^2}K^n\ ,
\end{equation}
where $\zeta<0$ is in agreement with~\eqref{A}, which indicates that $A$ decreases as $M$ increases [see Eq.~\eqref{effecmass} and~\eqref{deffin}]. Hence, for a given polytrope $\{K,\,n\}$, according to Eqs.~\eqref{efecprera},~\eqref{mimic2} and~\eqref{tolmanpressure} we find the pressure as
\begin{equation}
	\label{finalp}
	\tilde{p}_r(r)=p_r+{\cal P}_r=
	\frac{3\left(1+K^\Gamma\right)(R^2-r^2)}{\kappa\left(A^2+3\,R^2\right)\left(A^2+2r^2\right)}\ ,
\end{equation}
where we have used the condition in Eq.~\eqref{C}. On the other hand, the energy density is given by the expression in Eq.~\eqref{efecden}, where $\rho(r)$ is displayed in Eq.$  $~\eqref{tolmandensity} while the polytropic density takes the simple form
\begin{equation}
	{\cal E}=\left(\frac{K}{3\,\kappa^n}\right)\left[\frac{R^2}{A^2(A^2+3R^2)}\right]^\frac{n}{n+1}\left(3\,F_n+r\,F'_n\right)
	\ .
\end{equation}
The tangential pressure, given by Eq.~\eqref{efecpretan}, also has an analytical expression in terms of $F_n$ (which converge rapidly), but it is too large to display. As we see, our solution does not require any perturbative analysis. Fig.~\ref{fig1} shows the pressure in Eq.~\eqref{finalp} as a continuous function of the polytropic parameters $\{n,\,K\}$. We see that the effects are greater for the innermost layers, and are always proportional to $K$ and $1/n$. The same total effective pressure~\eqref{finalp} is displayed in Figs.~\ref{fig2} and~\ref{fig3}, now showing the effects of polytropes on stellar spheres explicitly. On the other hand, Fig.~\ref{figx} shows the pressure $\tilde{p}_r(r,K)$ and $\tilde{p}_r(r,n)$. Finally, the interaction between the polytope and the perfect fluid, which produces anisotropic consequences, is shown in Fig.~\ref{fig4}. We see that the interaction between both fluids increases significantly near the stellar surface, and in fact, there is a positive gradient of energy in the radial direction. This could be interpreted as the necessary work done by the polytrope to keep the perfect fluid within the stellar volume. We conclude by mentioning that the strong energy condition is satisfied in all regions inside the stellar distribution.
\section{Conclusions}
\label{con}
The study of relativistic fluids and their coexistence within self-gravitating systems is, in general, a complicated task to carry out. The reason for this lies in the complexity of Einstein's field equations, which introduces nonlinear effects that are difficult to handle, even for simplest cases such as static and spherically symmetric systems. Despite this intrinsic and ineluctable difficulty, in this work we have developed a simple, analytical and direct strategy to study the effects of polytropes on any other relativistic fluid, regardless of the nature of the latter.

As a direct application, we study the case of a perfect fluid coexisting with a polytrope characterized by the parameters $\{K,\,n\}$. To carry out the above, we use the well-known Tolman IV solution, which underlies in the limit $K\rightarrow\,0$, where all polytropic effects vanish. The total effective solution, formed by both fluids, is then analyzed, finding energy gradients $\Delta\,{\rm E}$ that increase in the radial direction. These gradients are maximum on the stellar surface $r=R$, as indicated in Fig.~\ref{fig4}, and are positive (negative) for the polytrope (perfect fluid). This indicates that the polytrope needs to give up energy to achieve a coexistence with the perfect fluid compatible with the exterior Schwarzschild solution. 

Finally, we want to point out that our solution satisfies the strong energy condition. However, it is necessary to carry out a more detailed study on its stability, and other questions that remain open, and that are beyond the goal of this work. For example, how much do our results depend on the chosen isotropic solution? How stable is the coexistence under radial perturbations? Likewise, the extension of this study to coexistence with other sources that are not necessarily isotropic remain open.

We want to conclude by emphasizing the direct impact of our approach on theories beyond Einstein, which can be described by a modified Einstein-Hilbert action as
\[
	\label{ngt}
	S_{\rm G}=S_{\rm EH}+S_{\rm X}=\int\left[\frac{(R-2\Lambda)}{2\kappa}+{\cal L}_{\rm M}+{\cal L}_{\rm X}\right]\sqrt{-g}\,d^4\,x\ ,
\]
where $R$ is the Ricci scalar, ${\cal L}_{\rm M}$ contains any matter fields appearing in the theory and ${\cal L}_{\rm X}$ the Lagrangian density of a new gravitational sector not described by general relativity, whose energy-momentum tensor is given by
\[
	\theta_{\mu\nu}=\frac{2}{\sqrt{-g}}\frac{\delta\,(\sqrt{-g}\,{\cal L}_{\rm X})}{\delta\,g^{\mu\nu}}=2\,\frac{\delta\,{\cal L}_{\rm X}}{\delta\,g^{\mu\nu}}-\,g_{\mu\nu}{\cal L}_{\rm X}\ .
\]
Therefore, following our scheme, we can always study the possible exchange of energy $\Delta{\rm E}$ between Einstein's gravity and any other gravitational sector not described by general relativity. 

\subsection*{Acknowledgments}
J.O.~is partially supported by ANID FONDECYT grant ${\rm N}^{\rm o}$ 1210041. EC is suported by Polygrant ${\rm N}^{\rm o}$  17459
%
%

\bibliography{references.bib}
\bibliographystyle{apsrev4-1.bst}
%
%
\end{document}